\documentclass[11pt,a4paper]{article}
\pdfoutput=1
\usepackage{jcappub}

\usepackage{graphicx}
\usepackage{amsmath, amssymb}
\usepackage{float}

\graphicspath{{figs/fracturing/},{figs/fracturing/visit/},{figs/fracturing/spectra/}}

\newcommand{\figref}[1]{figure~\ref{#1}}
\newcommand{\Figref}[1]{Figure~\ref{#1}}

\newcommand{\sgvonefield}{\url{www.cita.utoronto.ca/~jbraden/Movies/sg_v1_field.avi}}
\newcommand{\sgonerho}{\url{www.cita.utoronto.ca/~jbraden/Movies/sg_v1_rho.avi}}
\newcommand{\sgvonespec}{\url{www.cita.utoronto.ca/~jbraden/Movies/sg_v1_spec.avi}}

\newcommand{\sgvsqrtfield}{\url{www.cita.utoronto.ca/~jbraden/Movies/sg_vsqrt2_field.avi}}
\newcommand{\sgvsqrtrho}{\url{www.cita.utoronto.ca/~jbraden/Movies/sg_vsqrt2_rho.avi}}

\newcommand{\sgcascadefield}{\url{www.cita.utoronto.ca/~jbraden/Movies/cascade_field_wfluc.avi}}
\newcommand{\sgcascaderho}{\url{www.cita.utoronto.ca/~jbraden/Movies/cascade_contour_wfluc.avi}}

\newcommand{\lowvfield}{\url{www.cita.utoronto.ca/~jbraden/Movies/field_v0.05.mp4}}
\newcommand{\lowvrho}{\url{www.cita.utoronto.ca/~jbraden/Movies/econtour_v0.05.mp4}}
\newcommand{\lowvspec}{\url{www.cita.utoronto.ca/~jbraden/Movies/v0.05_spec.avi}}

\newcommand{\nondegenfield}{\url{www.cita.utoronto.ca/~jbraden/Movies/field_nondegen.avi}}
\newcommand{\nondegenrho}{\url{www.cita.utoronto.ca/~jbraden/Movies/econtour_nondegen.mp4}}
\newcommand{\nondegenspec}{\url{www.cita.utoronto.ca/~jbraden/Movies/nondegen_spec.avi}}

\newcommand{\onewallfield}{\url{www.cita.utoronto.ca/~jbraden/Movies/field_shapemode.avi}}
\newcommand{\onewallspec}{\url{www.cita.utoronto.ca/~jbraden/Movies/shapemode_spec.avi}}

\newcommand{\oscillonrho}{\url{www.cita.utoronto.ca/~jbraden/Movies/oscillon_rho.avi}}
\newcommand{\oscillonspec}{\url{www.cita.utoronto.ca/~jbraden/Movies/oscillon_spec.avi}}

\def\ie{i.e.}
\def\simord{\mathord{\sim}\,}
\def\laplacian{Laplacian}

\begin{document}

\title{Cosmic Bubble and Domain Wall Instabilities II: Fracturing of Colliding Walls}

\author[a,b,c]{Jonathan Braden,}
\affiliation[a]{CITA, University of Toronto, 60 St. George Street, Toronto, ON, M5S 3H8, Canada}
\affiliation[b]{Department of Physics, University of Toronto, 60 St. George Street, Toronto, ON, M5S 3H8, Canada}
\affiliation[c]{Department of Physics and Astronomy, University College London, London, WC1E 6BT, UK}
\emailAdd{j.braden@ucl.ac.uk}
\author[a]{J.\ Richard Bond}
\emailAdd{bond@cita.utoronto.ca}
\author[d]{and Laura Mersini-Houghton}
\affiliation[d]{Department of Physics and Astronomy, UNC Chapel Hill, NC 27599, USA}
\emailAdd{mersini@physics.unc.edu}

\date{\today}

\abstract{We study collisions between nearly planar domain walls including the effects of small initial nonplanar fluctuations.
These perturbations represent the small fluctuations that must exist in a quantum treatment of the problem.
In a previous paper, we demonstrated that at the linear level a subset of these fluctuations experience parametric amplification as a result of their coupling to the planar symmetric background.
Here we study the full three-dimensional nonlinear dynamics using lattice simulations, including both the early time regime when the fluctuations are well described by linear perturbation theory as well as the subsequent stage of fully nonlinear evolution.
We find that the nonplanar fluctuations have a dramatic effect on the overall evolution of the system.
Specifically, once these fluctuations begin to interact nonlinearly the split into a planar symmetric part of the field and the nonplanar fluctuations loses its utility.
At this point the colliding domain walls dissolve, with the endpoint of this being the creation of a population of oscillons in the collision region.
The original (nearly) planar symmetry has been completely destroyed at this point and an accurate study of the system requires the full three-dimensional simulation.}

\maketitle

\section{Introduction}
In this paper we extend the study begun in~\cite{ref:bbm1} of fluctuations around colliding parallel planar domain walls. While the previous paper~\cite{ref:bbm1} considered linearized fluctuations, here we examine the full nonlinear dynamics using lattice simulations. The results for nearly SO(2,1) bubble collisions are presented in the companion paper~\cite{ref:bbm3}.

Planar domain walls are a common ingredient in braneworld cosmology model building~\cite{Randall:1999ee,Rubakov:1983bb,Langlois:2002bb,Martin:2003yh,Bucher:2001it}, where they appear either as topological defects in a scalar field theory or as fundamental objects such as D-branes~\cite{Burgess:2001fx,Dvali:1998pa}.
Similarly, D-branes appear in various stringy inflation models and cyclic universe cosmologies~\cite{Khoury:2001wf}.
Since the presence of at least one wall (or brane) must be postulated in these setups,
it is natural to also consider the case when multiple walls are present.
These walls then evolve according to some dynamics, which may result in collisions.
In the general case these collisions occur in a wide variety of orientations, and the walls will have varying levels of bending and rippling.
However, braneworld setups often consider the case when collisions occur only between planar walls that are parallel to each other.
The planar approximation is also reasonable in other circumstances; for example collisions between nucleated bubbles that expand to several times their original size before colliding, or between approximately planar subsections of walls.

Since we study the full nonlinear dynamics between nearly planar extended objects, we need the high energy completion rather than just the effective theory for small fluctuations in the planar shape.
We focus on domain walls formed by the condensate of some scalar field, and thus we take the high energy completion of our theory to be a single field-scalar theory with canonical kinetic terms and symmetry breaking potential.
We refer to this field $\phi$ as the symmetry breaking scalar.

The case of interacting parallel planar walls in this class of theories has been considered by many authors, 
usually under the assumption that the nonlinear dynamics can be treated as exactly planar thus reducing the system to a single spatial dimension. 
This assumption is tenable in a \emph{classical} field theory.
However, as we noted in~\cite{ref:bbm1}, individual realizations of the quantum fluctuations break the planar symmetry once the theory is quantized.
The fluctuations can experience instabilities when the domain walls are allowed to collide with each other. 
These instabilities have important consequences for the full dynamics which we investigate in the remainder of this paper.

Several authors have explored the amplification of \emph{additional} fields coupled to the symmetry breaking scalar, including both the cases of additional fermionic fields~\cite{Gibbons:2006ge,Saffin:2007ja,Saffin:2007qa} and additional scalar fields~\cite{Takamizu:2004rq}.
However, these past studies assumed the background maintained planar symmetry and did not investigate the onset of nonlinearities amongst the fluctuations.
We instead take a more minimal approach and study the amplification of fluctuations in the symmetry breaking field $\phi$.
Since $\phi$ must exist to form the domain walls, these fluctuations are present in a consistent quantum treatment of the problem.
As anticipated in~\cite{ref:bbm1} we find that accounting for these fluctuations can drastically change the collision dynamics between the walls, in the process completely invalidating the original assumption of planar symmetry.
We refer to this situation as a breaking of the symmetry by fluctuations.
The process of symmetry breaking \emph{cannot} be captured by (1+1)-dimensional simulations such as those used in the exactly planar case.

The remainder of the paper is organized as follows.
In section~\ref{sec:models_kinks} we present our scalar field models and review  the domain wall solutions they support.  We also briefly review how these solutions interact in the limit of exact planar symmetry.
Section~\ref{sec:3d_wall_dynamics} constitutes our main results.
We use lattice simulations to study the full collision dynamics between domain wall-antiwall pairs for two different potentials and several choices of initial conditions.
A generic outcome of these collisions is the rapid amplification of nonplanar fluctuations, eventually leading to an inhomogeneous dissolution of the wall and antiwall.
While some of the energy is released into the bulk as radiation during the dissolution, some of it remains trapped in the collision region in the form of localized oscillating blobs of field called oscillons.
Motivated by the creation of oscillons from domain wall collisions, section~\ref{sec:oscillons} looks at some of their properties.
Finally, we provide a brief qualitative summary of the full collision dynamics in section~\ref{sec:summary_dynamics} and then conclude in section~\ref{sec:conclusion}.

\section{Review of (Early Time) Linear Fluctuation Dynamics}
\label{sec:models_kinks}
As in our study of linear fluctuations~\cite{ref:bbm1}, we consider two potentials that support kink solutions in one-dimension: the sine-Gordon model
\begin{equation}
  V(\phi) = \Lambda\left[1-\cos\left(\frac{\phi}{\phi_0}\right)\right]
  \label{eqn:potential_sg}
\end{equation}
and the double-well model
\begin{equation}
  V(\phi) = \frac{\lambda}{4}\left(\phi^2-\phi_0^2\right)^2 - \delta\lambda\phi_0^3\left(\phi-\phi_0\right) + V_0 \, .
  \label{eqn:potential_double_well}
\end{equation}
For the double-well, $\delta$ controls the difference between the false and true vacuum energies $\Delta\rho \approx 2\delta\lambda\phi_0^4$ and $V_0$ is a constant.
We restrict to Minkowski space, so $V_0$ will not play any role in the dynamics.
Unless explicitly indicated, we express the fields in units of $\phi_0$, spacetime coordinates in units of $m_{norm}^{-1}$, and energy densities in units of $m_{norm}^2\phi_0^2$ where $m_{norm}$ refers to the natural mass scale of the sine-Gordon or double well potential as appropriate (see~\eqref{eqn:sg_kink_ch3} and~\eqref{eqn:dw_kink_ch3}).
Both of these potentials have solutions (known as kinks and antikinks in one-dimension) interpolating between neighbouring minima of the potential.
For the sine-Gordon model and degenerate double-well ($\delta = 0$), they are given by
\begin{equation}
  \phi_{kink}^{SG} = 4\phi_0\tan^{-1}\left(e^{m_{SG}(x-x_0)}\right) \qquad m_{SG} = \sqrt{\Lambda}\phi_0^{-1}
  \label{eqn:sg_kink_ch3}
\end{equation}
and
\begin{equation}
  \phi_{kink}^{DW} = \phi_0\tanh\left(\frac{m(x-x_0)}{\sqrt{2}}\right) \qquad m=\sqrt{\lambda}\phi_0
  \label{eqn:dw_kink_ch3}
\end{equation}
respectively.  The antikinks are obtained via the replacement $(x-x_0) \to -(x-x_0)$.  
For the slightly asymmetric well ($\delta\ll 1$) stationary kink solutions no longer exist as the pressure differential between the false and true vacuum causes the kinks to accelerate.
In this case, we take the following approximate initial profile interpolating between the two minima
\begin{equation}
  \phi_{kink}^{DW} = \frac{\phi_{true}-\phi_{false}}{2}\tanh\left( \frac{m(x-x_0)}{\sqrt{2}} \right) + \frac{\phi_{true}+\phi_{false}}{2}
\end{equation}
where $\phi_{false}$ and $\phi_{true}$ are the locations of the false and true vacua repectively.
In one-dimension the energy of the sine-Gordon and double well kinks are 
\begin{equation}
  E_k^{SG} = 8m_{SG}\phi_0^2 \qquad and \qquad E_k^{DW}=\frac{2\sqrt{2}}{3}\sqrt{\lambda}\phi_0^3
\end{equation} 
respectively.
When the generalization of the kinks are embedded as domain walls in higher spatial dimensions, these energies become the surface tension of the wall.

In this paper we focus on collisions between a single kink-antikink pair, which together carry no net topological charge.
Since we are interested in the three-dimensional problem,
we extend the kink and antikink in the additional transverse spatial dimensions.
We refer to this setup as a wall-antiwall pair to distinguish it from the one-dimensional case.

Under the approximation of exact planar symmetry, the field obeys
\begin{equation}
  \frac{\partial^2\phi_{bg}}{\partial t^2} - \frac{\partial^2\phi_{bg}}{\partial x^2} + V'(\phi_{bg}) = 0
\end{equation}
with the initial condition
\begin{equation}
  \phi_{bg}(t=0) = \phi_{kink}(x-x_0) + \phi_{antikink}(x+x_0) + \phi_{\infty} \, .
  \label{eqn:phibg_init}
\end{equation}
We have aligned our axes so that the kink and antikink move along the x direction and collide at $x=0$.
The constant $\phi_{\infty}$ is chosen so that the field is sitting at the desired minimum of the potential at infinity.
For our purposes, the important aspect of these collisions is that they tend to produce oscillatory behaviour in the motion of the fields.
This oscillatory motion comes in three forms: repeated collisions between the kink and antikink, formation of localized pseudostable nearly periodic blobs of field, and for the double well the vibration of internal excitation modes of the individual kink and antikink.
When we consider the extension of the one-dimensional kinks to planar walls in higher spatial dimensions, small nonplanar fluctuations obey
\begin{equation}
  \partial_{tt}\delta\tilde{\phi}_{k_\perp} - \partial_{xx}\delta\tilde{\phi}_{k_\perp} + \left[k_\perp^2 + V''(\phi_{bg}(x,t)) \right]\delta\tilde{\phi}_{k_\perp} = 0
\end{equation}
where $\delta\tilde{\phi}_{k_\perp}$ is the 2D Fourier transform of the fluctuations in the additional orthogonal directions and $k_\perp$ is the transverse wavenumber.
The time-dependence of the one-dimensional kink-antikink solution induces a space- and time-dependent effective mass 
\begin{equation}
  m_{eff}^2(x,t) = k_\perp^2 + V''(\phi_{bg}(x,t))
\end{equation}
 for each transverse Fourier mode.
In~\cite{ref:bbm1} we performed a detailed analysis of the fluctuations accounting for this time-dependent effective mass.
We found that the oscillations in the background drive resonant instabilities in the fluctuations causing certain transverse wavenumbers $k_\perp$ to grow exponentially.
Eventually these fluctuations become sufficiently large that the assumption of linearity fails.
At this stage we have to solve the full nonlinear three-dimensional problem and waive any additional symmetry assumptions.
This full problem is the focus of the remainder of this paper.

\section{Nonlinear Dynamics of Planar Domain Walls with Non-Planar Fluctuations}
\label{sec:3d_wall_dynamics}
In this section we present the results for the full three-dimensional nonlinear field dynamics.
We only consider choices of the couplings
such that the fluctuations become highly excited while still in the linear regime.
Thus, the system transitions to the semiclassical wave limit before interactions between the fluctuations become important.
Invoking the standard assumption that the system remains in the semiclassical limit after the onset of strong nonlinearities amongst the fluctuations,
we can then use classical statistical simulations as an approximation to the full quantum evolution.
We use a high resolution numerical lattice code with second-order accurate/fourth-order isotropic finite-differencing stencils~\cite{Frolov:2008hy,Patra:2006} and sixth-order accurate Yoshida integrators for the time-evolution~\cite{Yoshida:1990,Huang:2011gf,Sainio:2012mw}.
Since we are interested in configurations that are spatially localized along the collision direction, we also implement absorbing boundary conditions along the longitudinal direction
\begin{equation}
  \left[\partial_t\phi - \partial_{x_\parallel}\phi \right|_{x_\parallel=0} = 0 \qquad \left[\partial_t\phi + \partial_{x_\parallel}\phi\right|_{x_\parallel=L_\parallel} = 0
\end{equation}
in order to remove energy released from the collision region~\cite{EngquistMajda:1997}.
We have chosen coordinates along the collision axis to range from $0$ to $L_\parallel$ and denoted the coordinate along this axis $x_\parallel$.
Quantum effects are incorporated through the initial conditions
\begin{align}
  \label{eqn:field_split_init}
  \phi_{init}({\bf x},t=0) = \phi_{bg}(x_\parallel,t=0) + \delta\phi({\bf x}) \\
  \notag \dot{\phi}_{init}({\bf x},t=0) = \dot{\phi}_{bg}(x_\parallel,t=0) + \delta\dot{\phi}({\bf x})
\end{align}
where $\phi_{bg}$ is the initial profile of the desired classical background field (here a pair of walls as in~\eqref{eqn:phibg_init}) 
and $\delta\phi$ and $\delta\dot{\phi}$ are realizations of random fields.
Although this approach cannot capture the final thermalization of the modes to the Bose-Einstein distribution,
it is capable of describing all forms of nonlinear mode-mode coupling, including mean-field like backreaction (as is included in the Hartree approximation), rescattering effects, and the development of nongaussian field statistics.
Since we focus on the dynamical regime well before the system reaches equilibrium, the lack of proper quantum thermalization is not a limitation.

An important ingredient in this framework is the statistics of the initial fluctuations.
In principle, the correlation functions of our realizations should match those of the quantum fluctuations they represent. 
Although we don't engage in a full study of the initial fluctuations here, 
as long as the exact initial state has a nonzero projection onto the linearly unstable Floquet modes the qualitative behaviour will be the same as in this paper.
For simplicity, we consider only two representative choices of initial fluctuations.  
The first is to take $\delta\phi$ and $\delta\dot{\phi}$ as homogeneous Gaussian random fields with spectra
\begin{align}
  \notag \langle|\delta\tilde{\phi}_k|^2\rangle \sim \frac{1}{2\sqrt{k^2+V''(\phi_{true})}} \\
  \langle|\delta\dot{\tilde{\phi}}_k|^2\rangle \sim \frac{\sqrt{k^2+V''(\phi_{true})}}{2} \, .
  \label{eqn:ic_bulkfluc_ch2}
\end{align}
These are the correct fluctuations if the background is homogeneous and sitting at its true vacuum minimum.
For the second choice we initialize each individual kink as
\begin{align}
  \label{eqn:ic_wallfluc_ch2}
  \phi_{init}({\bf x},t=0) &= \phi_{kink}(\gamma (x_\parallel + \delta x)) \\
  \dot{\phi}_{init}({\bf x},t=0) &= -\gamma (u+\delta u) \phi'_{kink}(\gamma(x_\parallel+\delta x)) \notag
\end{align}
where $\delta x(y,z)$ and $\delta u(y,z)$ are two-dimensional Gaussian random fields with spectra 
\begin{equation}
  \langle|\widetilde{\delta x}_{k_\perp}|^2\rangle \sim \frac{1}{2k_\perp \sigma_{kink}} \qquad \mathrm{and} \qquad \langle |\widetilde{\delta u}_{k_\perp}|^2\rangle \sim \frac{k_\perp}{2\sigma_{kink}} \, . 
\end{equation}
We have defined the transverse wavenumber squared $k_\perp^2 = k_y^2+k_z^2$ and the surface tension of the stationary kink $\sigma_{kink} = \int dx_\parallel (\partial_{x_\parallel}\phi)^2$.
The Lorentz contraction factor is given by $\gamma = (1-u^2)^{-1/2}$.
We only consider initial velocities with $u \ll 1$, so the inclusion of fluctuations does not lead to any superluminal wall propagation speeds.
In order to fix notation, we introduce an amplitude parameter $\mathcal{A}_b$ and initialize the fluctuations as
\begin{equation}
  \delta x(x_\perp) = \frac{\mathcal{A}_b}{L_\perp} \sum_{{\bf k}_\perp} \frac{\alpha_{\bf k_\perp}}{\sqrt{2k_\perp}} e^{i{\bf k}_\perp\cdot{\bf x}_\perp} \qquad  \delta u(x_\perp) = \frac{\mathcal{A}_b}{L_\perp} \sum_{{\bf k}_\perp} \beta_{\bf k_\perp}\sqrt{\frac{k_\perp}{2}} e^{i{\bf k}_\perp\cdot{\bf x}_\perp} \, .
\end{equation}
$\alpha_{\bf k_\perp}$ and $\beta_{\bf k_\perp}$ are complex Gaussian random deviates with variance $\langle|\alpha_{\bf k_\perp}|^2\rangle = 1 = \langle|\beta_{\bf k_\perp}|^2\rangle$, and $L_\perp$ is the side length of the box in the directions orthogonal to the collision.
The fluctuations described by~\eqref{eqn:ic_wallfluc_ch2} are local translations of the kinks, which are a subset of the full fluctuation content around the kink background.
In this second case we do not include the remaining bulk fluctuations (and transverse excitations of the shape mode in the case of the double-well).
From our linear analysis we know that for the case of well separated walls this set of localized fluctuations are precisely the modes which are most strongly amplified by the collision~\cite{ref:bbm1}.\footnote{Although all of the results we present here used one of these two choices, we also tested a variety of other initial conditions and obtained similar results.}
As well, with absorbing boundary conditions, the use of these localized initial conditions avoids an initial spurious loss of energy that occurs from absorption of bulk fluctuations by the boundaries.

For each sample collision we provide several plots illustrating different aspects of the evolution.
First, we slice the field along two orthogonal planes.
The first slice is parallel to the collision axis, providing a view of the effective one-dimensional dynamics at early times, the production of outgoing radiation, and the rippling of the walls as the transverse fluctuations are amplified.
The second slice is orthogonal to the collision axis and centered at either the collision point or the instantaneous location of one of the planar walls.  
This slice provides a full two-dimensional view of the development of the transverse instability.
As illustrated in the examples below, the second slice is especially useful to study the nonlinear evolution of the fluctuations.
The next set of figures are contour plots of the energy density $\rho \equiv -T^0_0=\frac{\dot{\phi}^2}{2} + \frac{(\nabla \phi)^2}{2} + V(\phi)$.
At early times these clearly show the locations of the two walls as well as the ripples that develop due to the linear fluctuations.
During the nonlinear stages the energy density plots provide a clear picture of how the system is evolving, including the emergence of structures localized in all three spatial directions.
Finally, to study the spectral content of the transverse fluctuations we include pseudocolor plots of the two-dimensional angle averaged power spectrum of $\rho$ for the transverse wavenumbers $k_\perp$ as a function of position along the collision axis.
Explicitly, we compute
\begin{equation}
  \mathcal{P}_{\rho}^{2d}(k_\perp,x_\parallel) \equiv \frac{L_\perp^2 k_\perp^2}{N_\perp^2} \left\langle|\tilde{\rho}^{2d}_{k_\perp}(x_\parallel)|^2\right\rangle_{\perp}
  \label{eqn:2d_power}
\end{equation}
where $\langle \cdot (x_\parallel)\rangle_{\perp}$ is an average over the plane orthogonal to the collision axis at position $x_\parallel$ along the collision axis. 
$N_\perp = N_yN_z$ is the number of lattice sizes in each of these planes.  
Our discrete Fourier transform convention is $\tilde{\rho}^{2d}_{k_\perp}(x_\parallel) = \sum_i e^{-i{\bf k}_{\perp,n}\cdot {\bf x}_{\perp,i}}\rho(x_{\perp,i},x_\parallel)$ with $x_\perp = (y,z)$ the coordinates in the directions orthogonal to the collision axis.
This gives us a very clear view of the spatial localization (along $x_\parallel$) of the amplified fluctuations as well as their typical transverse wavenumber.
In addition to the pseudocolor plot, we also show a slice of the spectrum at a fixed value of $x_\parallel$.
We pick this slice to either be through the center of one of the kinks or else through the center of the collision region.

Before presenting the results of our simulations, we briefly summarize the outcome of the collisions to orient the reader and unify the subsequent discussion.
While the precise evolution depends on the choice of potential and planar symmetric background $\phi_{bg}$, 
the qualitative details are essentially the same for every case we consider.
Initially, the two walls are well described by the planar ansatz.
Due to our choice of initial setup, the walls undergo multiple collisions or else capture each other to form an oscillating bound state.
In either case, the initially small planar symmetry breaking fluctuations experience rapid growth as described by linear theory~\cite{ref:bbm1}.
Once the fluctuations grow large enough they begin to interact nonlinearly.
In every case involving the dynamics of a wall-antiwall pair, we find that the stage of nonlinear interactions leads to a complete breakdown of the original planar symmetry.
This occurs by an inhomogeneous annihilation between the wall and antiwall.
The annihilation eventually results in the production of a population of oscillating blobs of field known as oscillons.
These oscillons are distributed homogeneously in the transverse directions to the collision, but are localized at the collision site along the collision axis.
Details and illustrations that clarify this picture are presented below.

\subsection{Sine-Gordon Potential}
We first analyze the sine-Gordon model.
We consider two distinct classes of background solutions.
The first class is the planar symmetric breathers indexed by the parameter $v$
\begin{equation}
  \phi_{breather} = 4\tan^{-1}\left(\frac{\cos(\gamma_v v t)}{v\cosh(\gamma_v x)}\right) \qquad \gamma_v \equiv (1+v^2)^{-1/2} \, .
\end{equation}
We restrict ourselves to $v \gtrsim 1$ so the background field configuration is a localized oscillating blob.
For the second class, we instead set the background solution to be a kink-antikink pair approaching each other with nonzero initial velocity.
Since the 1D sine-Gordon kinks are true solitons they preserve their shapes and velocities when they collide, acquiring only an overall phase shift due to the interaction.
Energy conservation thus dictates that after the collision the wall-antiwall pair again move off to infinity, at least in the absence of fluctuations.
When the system lives on the infinite interval we can use a Backlund-transformation to find a simple analytic form for the field
\begin{equation}
  \phi_{k\bar{k}} = 4\tan^{-1}\left(\frac{\sinh(\gamma ut)}{u\cosh(\gamma x)} \right)  \qquad \gamma \equiv (1-u^2)^{-1/2} \, ,
\end{equation}
which describes the passage of the kink and antikink through each other during the collision.
In this case the parameter $u$ represents the speed of the kink and antikink at infinity.
To allow for multiple collisions we make the collision direction periodic with a linear size less than the transverse directions.
As a result the wall and antiwall alternately collide in the middle of the simulation volume and at the boundaries.
Since the collision results in the passage of the wall and antiwall through each other, 
it is easy to see that each of these collisions will have the same ordering of the wall and antiwall relative to the collision site.  
Either we always have a kink approaching the collision from the right and an antikink approaching the collision from the left, or vice-versa.
\begin{figure}[!ht]
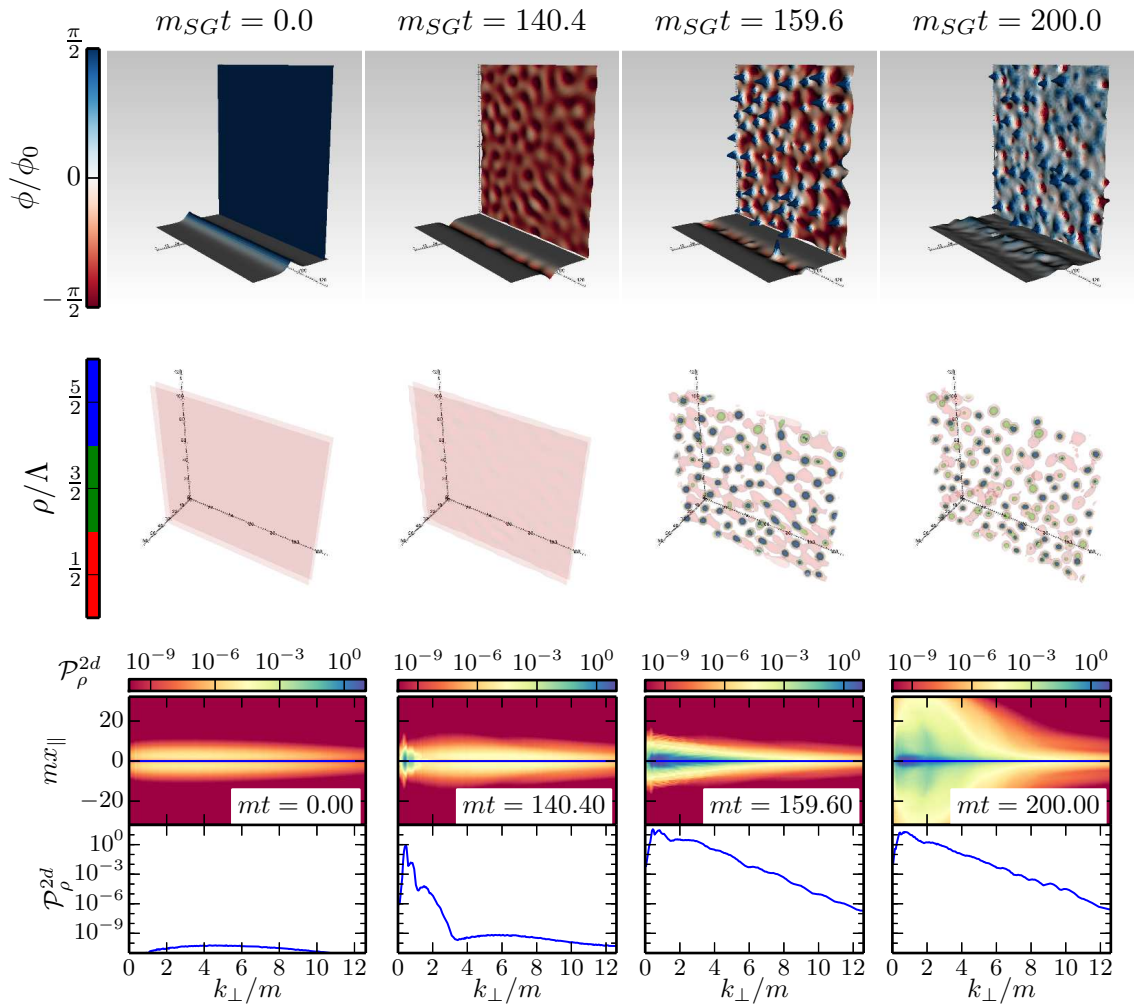

  \centering
    \includegraphics[width=0.99\linewidth]{{{sg_breather_vsqrt2_field_rho_multipanel}}} \\

    \includegraphics[height=2in]{{{sg_breather_vsqrt2_pspec_00000}}} 
    \includegraphics[height=2in]{{{sg_breather_vsqrt2_pspec_00702}}} 
    \includegraphics[height=2in]{{{sg_breather_vsqrt2_pspec_00798}}} 
    \includegraphics[height=2in]{{{sg_breather_vsqrt2_pspec_01000}}}
  \caption[Evolution of a $v=(\sqrt{2}-1)^{-1}$ sine-Gordon breather with small initial fluctuation around a planar background]{Evolution of a breather with $v=(\sqrt{2}-1)^{-1}$ showing the development of the instability in planar symmetry breaking fluctuations.  
    \emph{Top row:} The field sliced along a plane parallel to and orthogonal to the collision direction.  The orthogonal slice is taken through the center of the breather.  White shading corresponds to the field sitting at the origin and red and blue shading to the field displaced away from the origin.
    \emph{Middle row:} Contours of the energy density $\frac{\rho}{\Lambda} =\frac{\dot{\phi}^2}{2}+\frac{(\nabla\phi)^2}{2} + V(\phi)$.  
    \emph{Bottom row:} The dimensionless 2d power spectrum $\mathcal{P}_{\rho}^{2d}$ defined in~\eqref{eqn:2d_power}.  The top panel shows the spectrum as a function of $x_{\parallel}$ and $k_\perp$, while the bottom panel plots the spectrum along the slice through the center of the breather as indicated by the blue line in the top panel.  Animations of the field and energy density evolution can be found at \sgvsqrtfield\ and \sgvsqrtrho.}
  \label{fig:sg_vsqrt2}
\end{figure}

\subsubsection{Planar breathers with $v \leq 1$}
\label{sec:v1_breathers}
\begin{figure}[!ht]
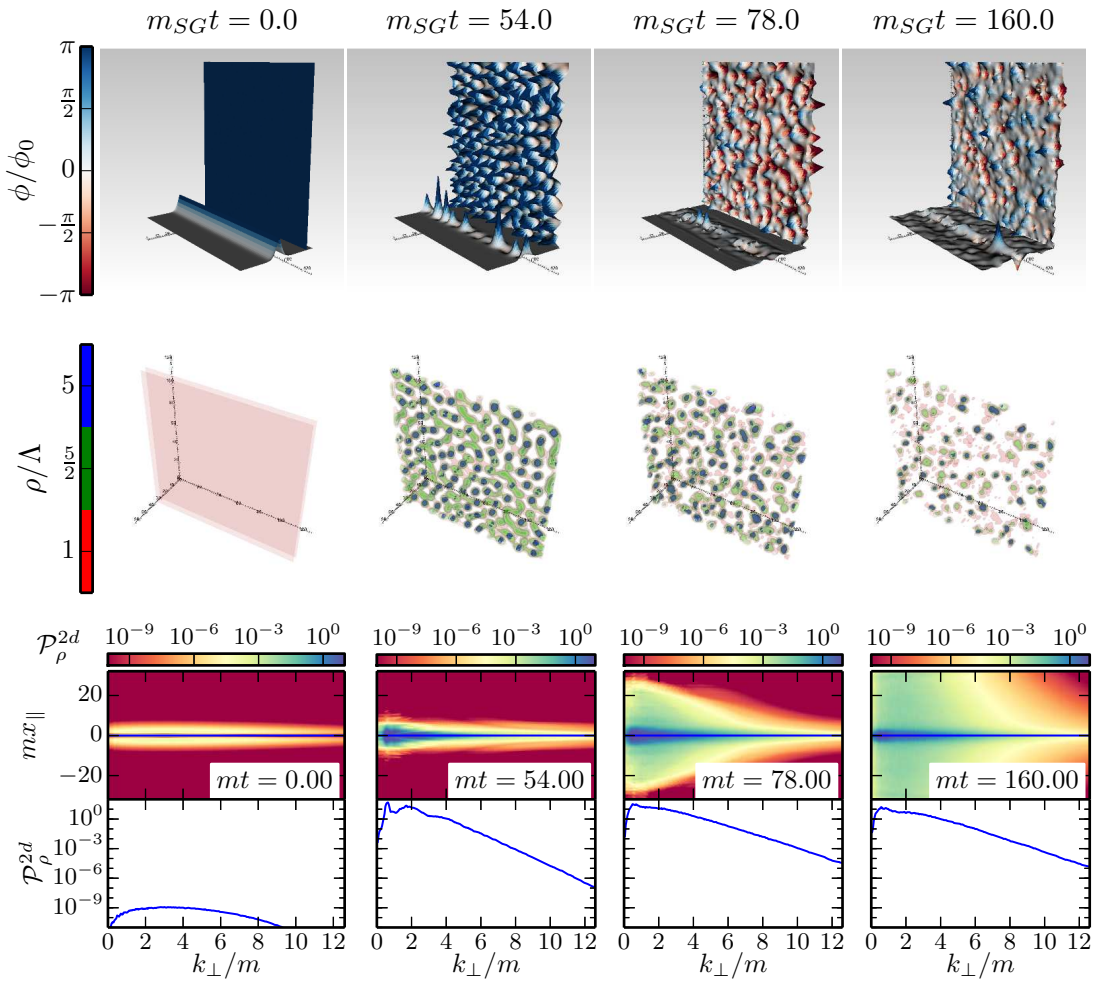

  \begin{center}
  \includegraphics[width=0.95\linewidth]{{{sg_breather_v1_field_rho_multipanel}}} \\
  \includegraphics[height=2in]{{{sg_breather_v1_pspec_00000}}}
  \includegraphics[height=2in]{{{sg_breather_v1_pspec_00135}}}
  \includegraphics[height=2in]{{{sg_breather_v1_pspec_00195}}}
  \includegraphics[height=2in]{{{sg_breather_v1_pspec_00400}}}
  \end{center}
  \caption[Evolution of the sine-Gordon breather with $v=1$ including small fluctuations around a planar background]{Evolution of the sine-Gordon breather with $v=1$ including small fluctuations around the planar symmetric solution. The choice of plots are the same as~\figref{fig:sg_vsqrt2}, although we have modified the color schemes slightly. Corresponding animations for the field, energy density and power spectrum evolution may be found at \sgvonefield, \sgonerho\ and \sgvonespec.}
  \label{fig:sg_v1}
\end{figure}

The evolution of planar symmetric breathers with small fluctuations are shown in~\figref{fig:sg_vsqrt2} and~\figref{fig:sg_v1} for the case $v=(\sqrt{2}-1)^{-1}$ and $v=1$ respectively.
At early times, the field is an oscillating blob localized along the collision axis with near planar symmetry in the transverse directions.
However, transverse fluctuations in a narrow band of $k_\perp$ are resonantly amplified by this oscillating background.
The fluctuations appear as ripples in the field profile and energy density contours, and as a growing peak in the dimensionless transverse power spectrum.
Since the planar breather is an exact solution for the one-dimensional sine-Gordon model, very little radiation is produced during this stage.
Of course, the linear growth of these fluctuations only continues until they begin to interact nonlinearly.
At this point the behaviour changes dramatically, and rescattering effects between the longitudinal and transverse modes destroys the clean separation between the planar background and the fluctuations.
The ripples in the breather from the transverse fluctuations become very large and pockets of field appear.
Outside of the pockets the field is near the origin, while in the interior it is displaced towards one of the two neighbouring vacua.
These pockets quickly condense into localized oscillating pseudostable blobs known as oscillons.
The oscillons are very long-lived, and they are held together by a competition between attractive forces from the potential and the dispersion induced by the \laplacian.
During this condensation a burst of radiation is released into the bulk, with additional radiation emitted by the subsequent slow decay of the oscillons.
In these cases, the characteristic transverse scale of the amplifed fluctuations is close the the final size of the oscillons, and the oscillons condense directly from the pockets of field formed by the linear instability.

\subsubsection{Kink-antikink collisions in a compactified extra dimension}
\begin{figure} [!ht]
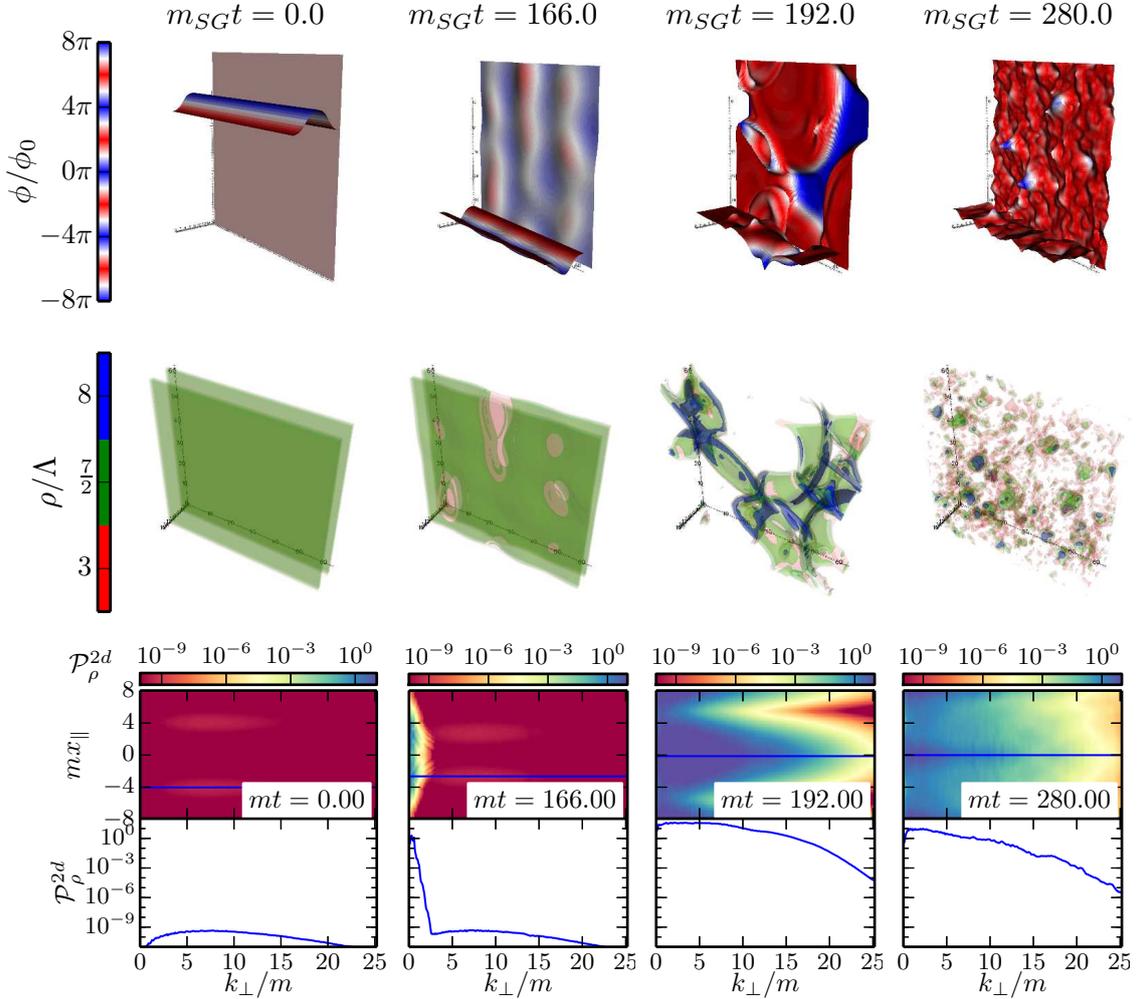

  \centering
  \includegraphics[width=0.99\linewidth]{{{sg_cascade_field_rho_multipanel}}} \\
    \includegraphics[height=2in]{{{sg_cascade_pspec_00000}}} 
    \includegraphics[height=2in]{{{sg_cascade_pspec_00415}}} 
    \includegraphics[height=2in]{{{sg_cascade_pspec_00480}}} 
    \includegraphics[height=2in]{{{sg_cascade_pspec_00700}}} 
  \caption[Evolution of repeated wall-antiwall collisions in the sine-Gordon model with periodic b.c.'s along the collision direction]{Evolution of repeated wall-antiwall collisions in the sine-Gordon model with periodic boundary conditions $\phi(x_\parallel+L_\parallel)=\phi(x_\parallel)$ along the collision direction.  \emph{Top Row}: Slices of the field parallel to and orthogonal to the collision axis.  \emph{Middle row}: Contours of the energy density.  \emph{Bottom row}: The evolution of the 2d angle averaged power spectrum $\mathcal{P}^{2d}_\rho$ defined in~\eqref{eqn:2d_power}.  The top panel plots $\mathcal{P}^{2d}_\rho$ as a function of transverse wavenumber $k_\perp$ and position along the collision axis $x_{\parallel}$.  The bottom panel plots the value along the blue line indicated in the top panel.  In all three rows, the data are taken at $m_{SG}t=0,166,192$, and $280$.  Corresponding animations can be found at \sgcascadefield\ and \sgcascaderho.}
  \label{fig:sg_kink_v0.2}
\end{figure}

Now let's consider the case of a colliding kink-antikink pair as illustrated in~\figref{fig:sg_kink_v0.2}.
The evolution is in many ways similar to the two breathers considered above.
Initially the field is well described by a colliding planar symmetric wall-antiwall pair.
This time the transverse fluctuations experience an inhomogeneous generalization of broad parametric resonance,
with the amplitude of fluctuations bound to the kink (and antikink) making a large jump at each collision.
The typical transverse wavelength of these fluctuations is closely related to the period of the background via $k_\perp \sim T_{collision}^{-1}$ where $T_{collision}$ is the time between collisions of the kinks.
Eventually the fluctuations become large enough that the next collision between the kink and antikink does not occur at the same time everywhere in space.
For our sample run, this inhomogeneous collision occurs in the middle of the domain.
From this point, the evolution deviates from the $v \geq 1$ breathers considered in section~\ref{sec:v1_breathers}.
Instead of pockets of field immediately condensing into oscillons, punctures form in the walls.
These punctures quickly turn into a collection of tubes threading the wall-antiwall pair.
The tubes expand in radius and eventually collide with each other, forming a planar network of fat filaments at the collision site.
When viewing a single (two-dimensional) orthogonal slice of the field taken through the center of the collision, the process is very reminiscent of bubble nucleation and expansion during a first-order phase transition.
The filaments then fracture into oscillons.
In the particular run shown in this paper, an additional piece of interesting dynamics also occurs.
During the final collision some segments of the walls are not captured at the origin but instead pass through the center of the domain and collide one final time at the periodic boundary.
Thus, a similar fracturing process occurs at the periodic boundary as well, and oscillons are produced in two distinct planes corresponding to the two collision sites of the background motion.
As with the $v \geq 1$ breathers, the final outcome of the collisions is completely different than for an exactly planar solution.
In the exactly planar case, the kink and antikink continue to collide with each other indefinitely (since they are solitons).
Therefore, given enough time the field can traverse an arbitrary distance in field space.
However, when nonplanar fluctuations are included the fields only traverse a finite distance in field space before the amplification of the fluctuations leads to the dissolution of the walls.

For this study, we have taken the sine-Gordon model to provide the potential for a single-field scalar model and have not imposed any additional identifications on the field.
However, one could imagine that $\phi$ is instead an angular degree of freedom in some two-field model with the radial degree of freedom effectively trapped at the minimum.  
In this case, we may expect the fracturing process to result in the excitation of the radial degree of freedom and possibly the production of global strings.
We do not explore this possibility here, although it could provide interesting phenomenology in models based on small compactified extra dimensions.
As well, a similar setup involving multiple collisions between a brane and antibrane is used in unwinding inflation~\cite{D'Amico:2012sz,D'Amico:2012ji,D'Amico:2014psa}.  Although the UV theory of a brane and a scalar domain wall are different, we expect that similar dynamics will also play a role in unwinding inflation.

\subsection{Double-Well Potential}
Thus far we demonstrated in the sine-Gordon model that the inclusion of small initial nonplanar fluctuations around colliding planar domain walls can have a drastic effect on the field evolution, ultimately leading to a complete breakdown of the initial near planar symmetry of the fields.
However, the sine-Gordon model in $1+1$-dimensions (\ie\ the planar limit) is integrable and thus rather special.
We now demonstrate that similar conclusions hold for the double-well potential~\eqref{eqn:potential_double_well}.

\subsubsection{Low Incident Velocity Collision in Symmetric Double Well}
\begin{figure}
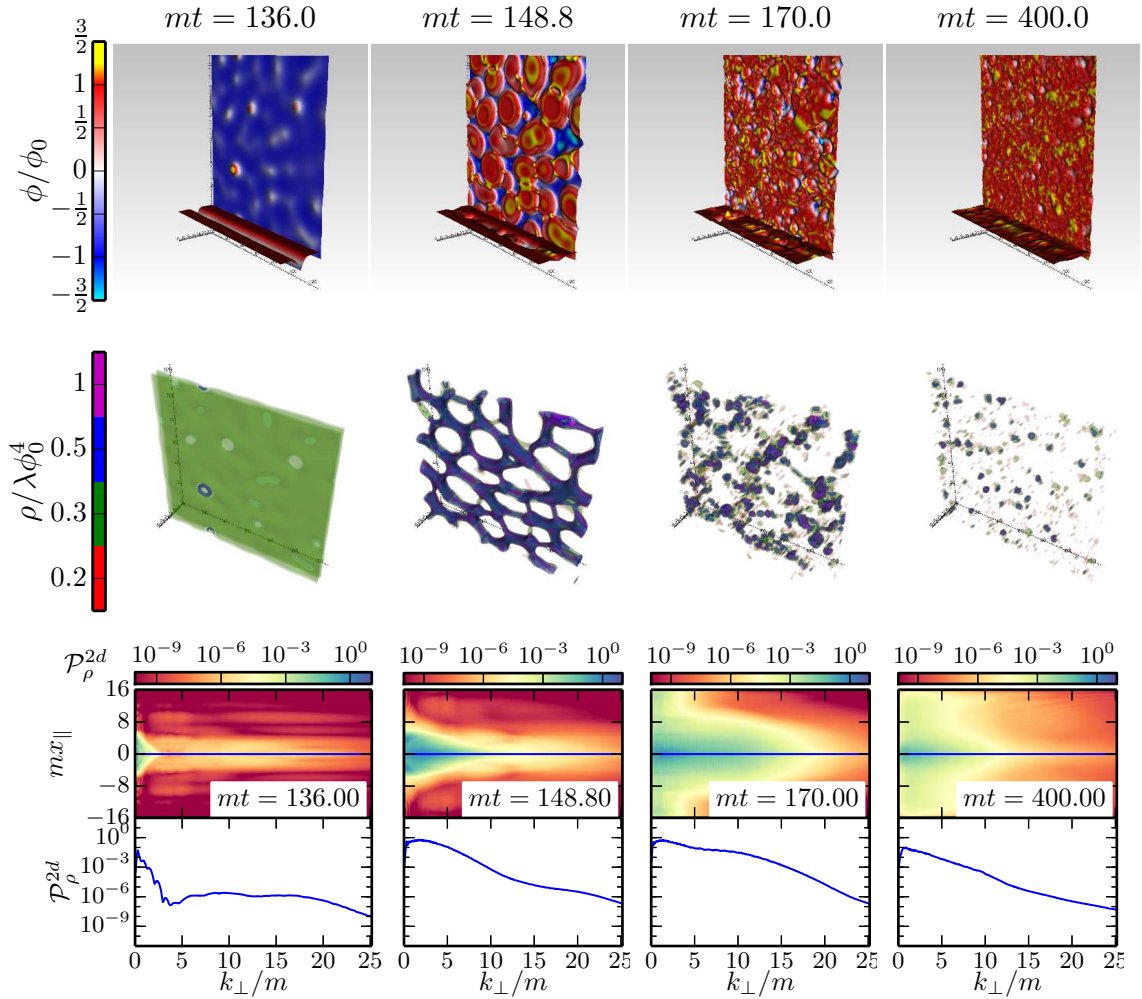

  \centering
  \includegraphics[width=0.99\linewidth]{{{v0.05_field_rho_multipanel}}}\\
  \includegraphics[height=2in]{{{v0.05_pspec_00340}}}
  \includegraphics[height=2in]{{{v0.05_pspec_00372}}}
  \includegraphics[height=2in]{{{v0.05_pspec_00425}}}
  \includegraphics[height=2in]{{{v0.05_pspec_01000}}}
  \caption[Various aspects of the time evolution of two colliding domain walls with initial speeds $u=0.05$]{Several snapshots demonstrating various aspects of the time evolution of two colliding domain walls in the symmetric double-well with initial speeds $u=0.05$.  From the top row to bottom: (a) the field distribution taken on a slice through the center of the collision ($mx_\parallel=16$), (b) contours of the energy density, (c) the dimensionless 2d angular averaged power spectrum for the energy density as a function of position along the collision direction.  We align our coordinates so that the collision occurs along the x-direction.  The simulation parameters were $mdx=0.125$, $mL_{\parallel}=64$, $mL_{\perp}=256$, and $mdt=0.025$. Absorbing boundary conditions were used at $mx_{\parallel}=0,64$.  For this initial speed, the energy density at the center of the wall is $\rho/\lambda\phi_0^4 = 0.5/(1-u^2)^{1/2} \approx 0.50125$.  A description of the dynamics is given in the main text.  An animation of the field is available at \lowvfield, the evolving energy density at \lowvrho, and the power spectrum at \lowvspec.}
  \label{fig:v0.05}
\end{figure}

\begin{figure}
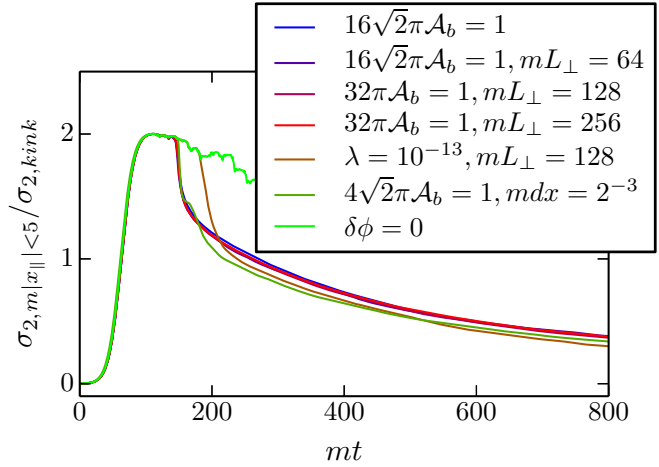

  \begin{center}
    \includegraphics[width=0.6\linewidth]{{{eslab_v0.05_new}}}
  \end{center}
  \caption[Energy within a slab of width $mL_{slab}=10$ centered on the initial collision of the two walls for $u=0.05$.]{Energy within a slab of width $mL_{slab}=10$ centered on the initial collision of the two walls for $u=0.05$.  We have plotted the result for both bulk fluctuations and transverse fluctuations in the wall's location, as well as for a range of box sizes and grid spacings.  Also shown for comparison is the result if no fluctuations are included.  Unless indicated in the legend, we used $mdx=0.25$, $mL_\perp=128$ and $mL_\parallel=64$.  If a value of $\mathcal{A}_b$ is listed in the legend we used initial conditions~\eqref{eqn:ic_wallfluc_ch2}, while if $\lambda$ is listed we used~\eqref{eqn:ic_bulkfluc_ch2}.  The oscillations in the planar solution are due to the slab being slightly smaller that the size of the region occupied by the oscillating blob.}
  \label{fig:eslab_v0.05}
\end{figure}

As a first example in the symmetric double-well, consider the case of low incident speed $u=0.05$ illustrated in~\figref{fig:v0.05}.
Initially, the fluctuations are small and the system is well described by the background configuration of two planar walls.
The walls then move towards each other and first collide at $mt \approx 110$.
After this initial collision the walls never become well separated and they appear as a localized oscillating blob that is similar to the $v \geq 1$ sine-Gordon breathers above.
A small amount of planar radiation is released during these oscillations, an effect which is properly captured by the symmetry reduced ($1+1$)-d dynamics as seen in~\figref{fig:eslab_v0.05}.
Far more importantly, a range of transverse fluctuations grow exponentially in the background of the oscillating planar symmetric ``blob''.
As these fluctuations grow, they appear as bumps and ripples perturbing the planar symmetry, which are evident in the energy density contours.
Eventually, these bumps become large enough that several sections of the planar blob pinch off, forming punctures of true vacuum.
As a result, the region where the field is displaced from the minimum is threaded by tubes where the field is near the true vacuum.
From this point on, the evolution is very similar to the sine-Gordon kink-antikink collisions in a compactified extra dimension.
These tubes expand and eventually coalesce, leading to a network of fat filamets with the field near the false vacuum in the interior.
This network is contained within a planar region of width $\simord m^{-1}$ along the collision axis and extends indefinitely in the directions orthogonal to collision (due to the original planar symmetry).
This process is illustrated in the first, second and third columns of~\figref{fig:v0.05}.
In Fourier space, the developing network of filaments manifests itself as a rapidly growing tail of fluctuation power that extends to $k_\perp \sim 15m$.
As the final step in the process, the filaments fracture into localized blobs of field --- the oscillons of the double-well potential.
Thus, exactly as in the sine-Gordon model, a population of oscillons is produced in the collision region as the endpoint of the dynamical amplification of the symmetry breaking fluctuations.

An important quantity is the amount of energy that escapes from the collision region as scalar radiation versus the amount that remains stored in oscillons.
In~\figref{fig:eslab_v0.05} we plot the energy density per unit transverse area
\begin{equation}
  \sigma_2 = \frac{1}{A_\perp}\int d^2x_\perp \int_{-L_\parallel/2}^{L_\parallel/2}dx_\parallel \rho  \qquad A_\perp = \int d^2x_\perp
  \label{eqn:2d_energy_density}
\end{equation}
remaining in the simulation box as a function of time.
The production and longevity of the oscillons prevents a complete release of the energy originally stored in the domain walls into the bulk.
However, relative to the case of exact planar symmetry much more of the energy escapes.
Of course, this occurs because in the planar limit the wall and antiwall don't immediately annihilate each other, but instead form a long-lived bound state.
The planar bound state is the one-dimensional version of the oscillon for this potential, and since it is very long lived the energy release in the planar case is slow.

\subsubsection{Interactions in a resonant escape band}
\begin{figure}
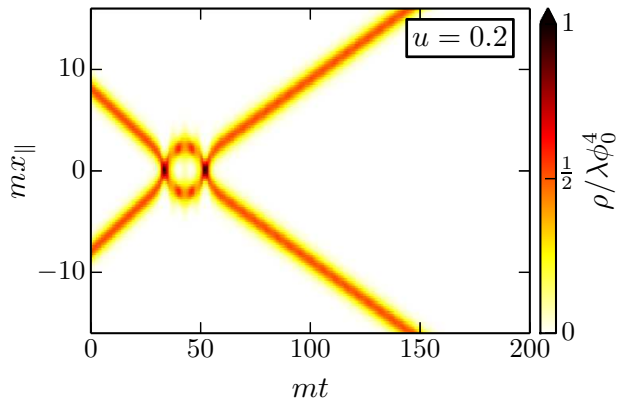

  \centering
  \includegraphics[width=0.6\linewidth]{{{1d_collision_rho_v0.2}}}
  \caption{Evolution of the energy density $\frac{\rho}{\lambda\phi_0^4}$ for a planar symmetric collision in the symmetric double-well with initial kink and antikink speeds given by $u=0.2$.  The location of the the kink and antikink are clearly visible as peaks in the energy density.  For this choice of initial speed, the wall-antiwall pair bounce off of each other twice before escaping back to infinity.}
  \label{fig:v0.2_bgrho}
\end{figure}

As an example of an even more dramatic effect induced by the breaking of planar symmetry, let's now choose an incident velocity of $u=0.2$.
In the planar case, the walls bounce twice before escaping back to infinity, which is illustrated in \figref{fig:v0.2_bgrho} and the top right panel of figure 3 of~\cite{ref:bbm1}.
During the first collision the planar shape mode is excited, draining energy from the kinetic motion of the (planar) walls.
In the second collision, the nonlinear field interactions transfer energy from the excited shape mode back into translational energy of the (planar) walls.
This transfer process clearly requires a tuning between the oscillation period of the shape mode and the time between collisions for the two walls.
If this tuning is disrupted or energy is drained from the planar oscillations of the shape mode, then rather than escape the walls will capture each other.

\begin{figure}
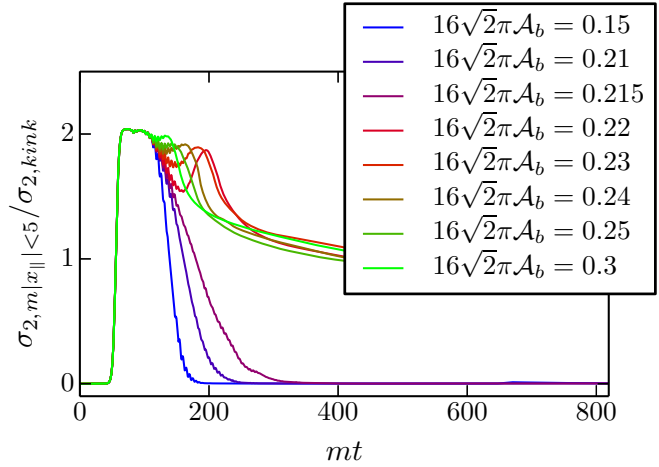

  \centering
  \includegraphics[width=0.6\linewidth]{{{eslab_v0.2_varyamp}}}
  \caption[Energy density within a slab of width $mL_{slab}=10$ for initial speed $u=0.2$ and various initial fluctuation amplitudes]{Energy density within a slab of width $mL_{slab}=10$ for initial speed $u=0.2$ and various initial fluctuation amplitudes $\mathcal{A}_b$ for the 2d Gaussian random field corresponding to the local translations in the wall position.  
For the simulations we used a box of size $L_\parallel=64$, $L_\perp=128$ and $dx=0.25$, and we used the initial fluctuations~\eqref{eqn:ic_wallfluc_ch2} with $\delta u=0$.}
  \label{fig:eslab_v0.2}
\end{figure}

When we include the transverse fluctuations, they experience a nonadiabatic kick at each collision.
As well, since the (homogeneous) shape mode is excited during the first collision, further pumping occurs while the walls are separated.
The energy required to amplify the fluctuations must be drained from the kinetic energy of the planar background and the oscillations of the planar shape mode.
Since the initial stage of fluctuation amplification is a linear effect, the size of this backreaction on the planar background increases with the initial amplitude of the fluctuations.
Therefore, for sufficiently large initial amplitudes the resulting backreaction will prevent the walls from escaping back to infinity.
In this case the qualitative behaviour of the system changes; instead of escaping the walls capture each other and fracture into oscillons.
To illustrate this \figref{fig:eslab_v0.2} shows the energy per unit area~\eqref{eqn:2d_energy_density}
within a slab of width $mL_{slab} =10$ centered on the collision for a range of initial fluctuation amplitudes.
For an isolated kink moving at speed $u$ with $L=\infty$ this energy is $\frac{\sigma_2}{\sqrt{\lambda}\phi_0^3} = \frac{2\sqrt{2}}{3\sqrt{1-u^2}}$.
Two distinct behaviours are visible: either $\rho$ drops abruptly to zero by $mt\sim 200$ corresponding to the case when the two walls escape back to infinity, 
or else $\rho$ slowly decays for $mt\gtrsim 200$ corresponding to the case when the walls capture and subsequently fracture into oscillons.
The transition between these two behaviours occurs as we increase the initial amplitude of the fluctuations.
We only included fluctuations associated with the translation mode of the walls~\eqref{eqn:ic_wallfluc_ch2} and also set $\delta u=0$.
Hence, this underestimates the true amount of backreaction on the walls and is only an illustration of the dramatic effect that fluctuations can have.
We have not carefully explored the space of initial conditions, so more complicated behaviour such as three bounces before escaping or annihilating may also be possible for finely tuned fluctuations.

\subsection{Asymmetric Double-Well Potential}
\label{sec:asymmetric}
Finally, consider collisions in the asymmetric double well with $\delta = 1/30$.
As a concrete example, we give the walls an initial separation $2mr_{init} = 16$ with initial speeds $u=0$.
The vacuum energy difference between the two wells causes the walls to accelerate and they collide with an initial speed $u^2=\frac{\mu(\mu+2)}{(1+\mu)^2} \sim 0.6$.
We've defined $\mu = r_{init}\Delta\rho/\sigma_{kink} \sim 0.57$ in terms of the energy difference between the two wells $\Delta\rho\sim 2\delta\lambda\phi_0^4$ and the suface tension of the wall $\sigma_{kink}=\frac{2\sqrt{2}}{3}\sqrt{\lambda}\phi_0^3$.
Because their relative velocity at collision is much larger than the cases considered above, the walls separate much further from each other between collisions.
However, the pressure induced by the nondegeneracy of the two vacua eventually turns the walls around and prevents them from escaping back to infinity.

\begin{figure}[ht]
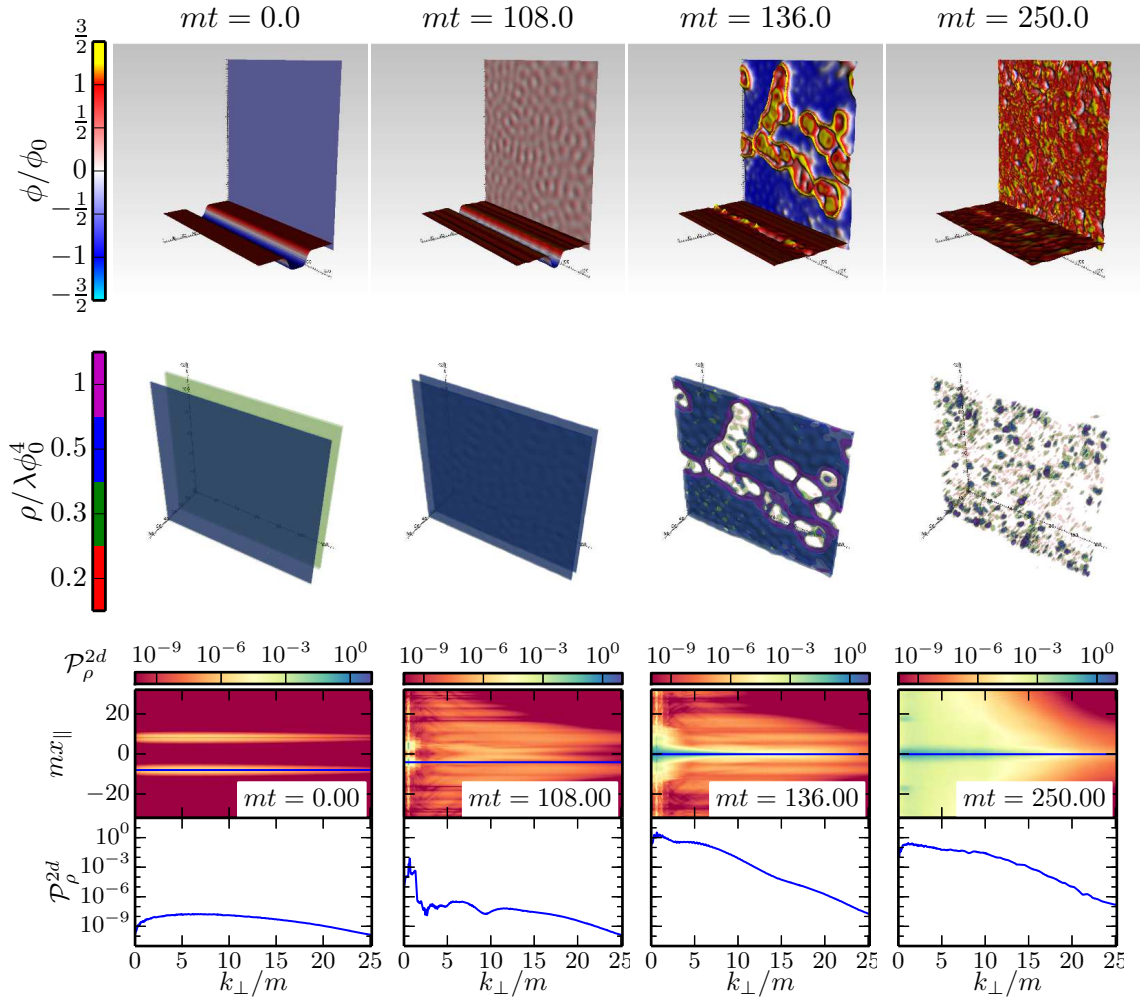

  \centering
    \includegraphics[width=0.99\linewidth]{{{del1o30_field_rho_multipanel}}}\\
    \includegraphics[height=2in]{{{del1o30_pspec_00000}}} 
    \includegraphics[height=2in]{{{del1o30_pspec_00270}}} 
    \includegraphics[height=2in]{{{del1o30_pspec_00340}}} 
    \includegraphics[height=2in]{{{del1o30_pspec_00625}}} 
  \caption[Dynamics of a colliding wall-antiwall pair in the asymmetric double well]{Collision dynamics in the asymmetric well with $\delta = \frac{1}{30}$ during several different stages of evolution.  
    The choice of illustrations match those in~\figref{fig:sg_vsqrt2}.
    From left to right, we have snapshots of the initial setup ($mt=0$), just prior to the final collision ($mt=108$), during the fracturing process ($mt=136$) and showing the final population of oscillons ($mt=250$).
  The blue line in the pseudocolor plot of the 2d power spectrum indicates both the position of the orthogonal slice of the field in the top row, as well as the slice along which the 2d spectrum is plotted (\emph{bottom panels, bottom row}).
Animations of the field evolution can be found at \nondegenfield, the energy density evolution at \nondegenrho, and the power spectrum evolution at \nondegenspec.}
  \label{fig:nondegen}
\end{figure}

The resulting behaviour is illustrated in~\figref{fig:nondegen}. 
The left column shows the initial well separated walls with small fluctuations.
The walls accelerate towards each other and bounce multiple times.
At each bounce a band of transverse fluctuations are excited.
As well, planar symmetric shape modes are excited by the collision.
While the walls are separated between collisions, the shape mode further excites transverse modes of the field.
By the time of the final collision, these transverse fluctuations have become quite large.
They are visible in the second column of~\figref{fig:nondegen} as bumps in the energy density contours, bumps in the field profile, and as peaks in the power spectrum.
The nearly planar radiation emitted during the early stages of the evolution is also visible in the field profile.

Aside from the pumping of fluctuations by the planar shape mode becoming a distinct process, 
the biggest difference between this case and the $u=0.05$ case in the symmetric well is the inhomogeneity of the final disintegration of the walls.
In the case we have illustrated, fluctuations with transverse wavelengths much larger than the typical size of the bumps in the walls and the oscillons that eventually form were excited.
As a result, clusters of true vacuum tubes threading the walls appear, as opposed to the regular honeycomb like structure from the $u=0.05$ case.
This can be seen in the third column of~\figref{fig:nondegen}.
Once again the walls become threaded by tubes where the field is near the true vacuum.
These tubes then expand producing a network of filaments with the field trapped near the false vacuum in the interior, just as in the $u=0.05$ case.
Subsequently these filaments fracture and a collection of oscillons is formed.

For the case illustrated here, the walls collide five times with the fifth collision resulting in the breakup of the walls and ultimately the production of oscillons.
Whether or not a long-wavelength mode was excited in an individual collision depends on the particular realization of the initial fluctuations used in the simulation.
An interesting feature of this particular run was the ejection of a pair of oscillons into the bulk during the breakup of the filamentary network.
These are visible as two isolated peaks at $mx_\parallel \sim 15,50$ in the transverse power spectrum in the right column of~\figref{fig:nondegen}.
The ejection of oscillons into the bulk was sensitive to the particular realization of the fluctuations.

\subsection{Growth of fluctuations from planar shape mode}
Thus far we have considered the three-dimensional dynamics of colliding wall-antiwall pairs (the higher dimensional equivalent of a kink-antikink pair in one-dimension) including small planar symmetry breaking fluctuations.
Among the cases we considered were the oscillations of tightly bound wall-antiwall pairs (such as the $v \geq 1$ sine-Gordon breathers) and repeated collisions between weakly bound pairs (such as the asymmetric double well).
In the latter case, the internal dynamics (specifically planar symmetric excitations of the shape mode) of each individual wall plays an important role both in the evolution of the background planar solutions as well as the growth of fluctuations.
This is especially true for the double-well potential at large collision velocities where the walls either escape back to infinity (symmetric well) or become well separated from each other between collisions (asymmetric well).
To isolate the effects of the internal wall dynamics, we now study a single domain wall at rest in the symmetric double well with an initial width different from that of the static kink solution
\begin{equation}
  \phi_{init} = \phi_0\tanh\left( \frac{m(x-x_0)}{mw} \right) + \delta\phi_{fluc} \, .
\end{equation}
The initial width of the wall is $mw \neq \sqrt{2}$ and as before $\delta\phi_{fluc}$ is a realization of a random field.
This choice of initial background mimics the excitation of a planar symmetric shape mode.

\begin{figure}[t]
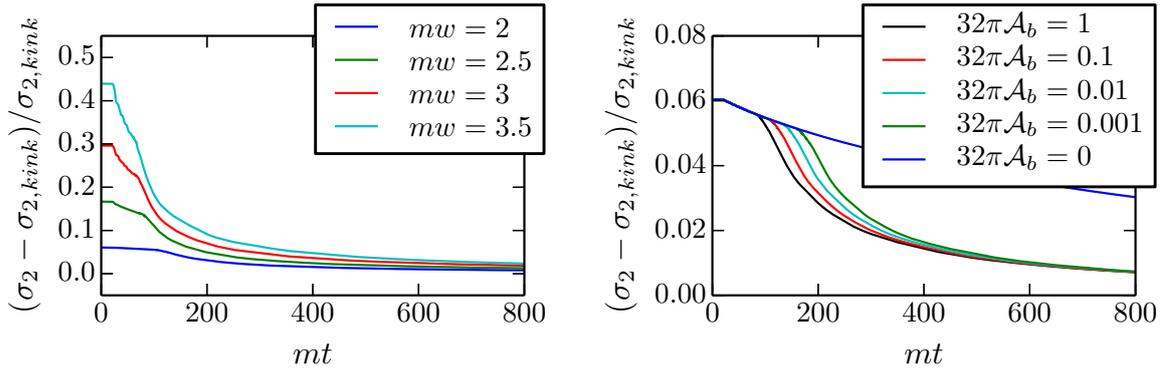

  \begin{center}
    \includegraphics[width=0.48\linewidth]{{{eslab_1wall_varywidth}}}
    \hfill
    \includegraphics[width=0.48\linewidth]{{{eslab_1wall_varyamp}}}
  \end{center}
  \caption[Transverse energy density within the simulation volume for an initially excited planar shape mode]{Time dependence of the excess energy density within the simulation volume for a box with lengths parallel and orthogonal to the collision given by $mL_{\parallel}=32$ and $mL_\perp = 128$ respectively. Absorbing boundary conditions were used in the collision direction and the grid spacing was $mdx = 0.25$.  \emph{Left:} The result for several choices of the initial wall width $w$.  \emph{Right:} The result for several choices of initial fluctuations and $mw=2$.  For comparison we also plot the excess energy when no initial fluctuations are present.  The effect of the fluctuations is clearly visible as a sharp change in the rate of energy loss relative to the planar case.}
  \label{fig:eslab_wall}
\end{figure}

Just as for collisions, the presence of transverse fluctuations radically modifies the behaviour of the field.
A very explicit demonstration of this comes from studying the energy contained within our simulation volume as shown in~\figref{fig:eslab_wall}.
Once again, initially the energy decreases via emission of planar symmetric radiation that can be captured via $1+1$-dimensional simulations.
However, as the shape mode oscillates it pumps transverse fluctuations, eventually leading to the onset of strong nonlinearities amongst the symmetry breaking fluctuations.
At this point energy is released much more rapidly and approaches the energy of a single isolated domain wall. 
As expected, the timing of this transition depends on the initial amplitude and spectrum of the fluctuations, but the subsequent
behaviour is rather insensitive to these details.

\begin{figure}[t]
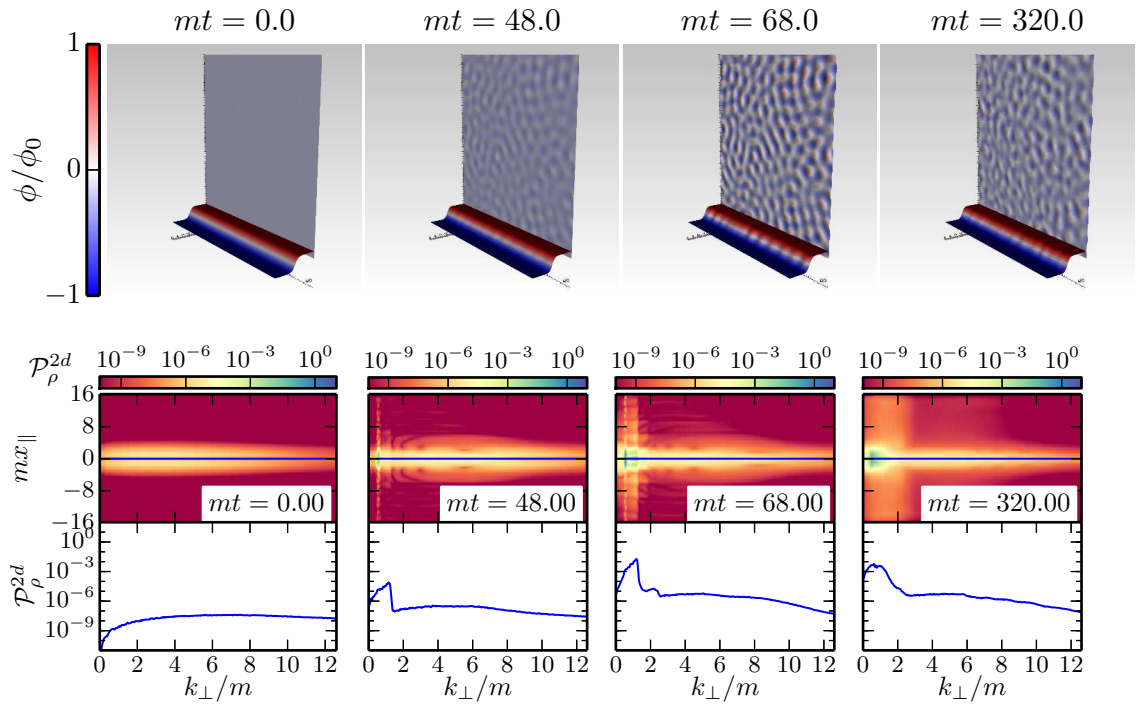

  \includegraphics[width=0.99\linewidth]{{{shapemode_fieldslice_multipanel}}}\\
    \includegraphics[height=2in]{{{shapemode_pspec_00000}}}
    \includegraphics[height=2in]{{{shapemode_pspec_00060}}}
    \includegraphics[height=2in]{{{shapemode_pspec_00085}}}
    \includegraphics[height=2in]{{{shapemode_pspec_00400}}} 
  \caption[Evolution of wall with excited planar shape mode and small initial fluctuations]{Several snapshots of the growth of transverse fluctuations due to excitation of the planar shape mode of the wall.  \emph{Top row:} Slice of the field $\phi$ parallel to the collision axis and orthogonal to the collision axis and centered on the middle of the wall.  \emph{Bottom row:} Two-dimensional power spectrum $\mathcal{P}_\rho^{2d}$ defined in~\eqref{eqn:2d_power}.  An animation of the field evolution can be found at \onewallfield\ and the power spectrum at \onewallspec.}
  \label{fig:wall_snapshots}
\end{figure}

To shed further light on the field dynamics and the origin of the decay in excess energy bound to the domain wall,~\figref{fig:wall_snapshots} shows several snapshots of the wall as it passes through the nonlinear phase.
Initially the field is nearly homogeneous and looks like a tanh function with a vibrating width.
This means that the well binding localized fluctuations to the wall has an oscillating width (and shape).
These oscillations pump a narrow band of transverse fluctuations as expected from linear perturbation theory.
As these fluctuations grow, a speckled pattern emerges in the field distribution taken along a slice through the middle of the wall.
This pattern is superimposed on the overall oscillation of the wall.
Eventually, the transverse oscillations enter the nonlinear regime, leading to a slight broadening of the spectral peak and emission of radiation from the wall.
The planar oscillation of the wall disappears and bound transverse lumps of field appear along the wall as seen in the third and fourth columns.
These transverse lumps  persist for times $mt \gtrsim 800$.
We can now understand the residual energy relative to an unexcited domain wall seen in~\figref{fig:eslab_wall} as being due to this population of ``wall lumps''.
Comparing the second column of~\figref{fig:nondegen} with the second column of~\figref{fig:wall_snapshots}, we can see that the pumping of fluctuations by the shape mode is indeed an important effect for collisions in the asymmetric well.

\section{Oscillons as a Long-Lived Intermediate State}
\label{sec:oscillons}
By solving for the full nonlinear evolution of wall-antiwall collisions, 
we found that the initial rapid amplification of linear fluctuations is followed by a very short stage during which the walls dissolve.
At the end of this stage, we are left with a population of quasi-stable localized oscillating blobs of field distributed in a narrow plane orthogonal to the collision axis.
These field structures are known as oscillons and arise from a dynamical balancing act between the dispersive effect of the \laplacian\ and the attractive force from the potential.
Oscillons were first discovered by Bogolyubsky and Makhankov~\cite{Bogolyubsky:1976yu} and then rediscovered by Gleiser~\cite{Gleiser:1993pt}.
There is a large literature devoted to their properties~\cite{Fodor:2006zs,Fodor:2008es,Fodor:2009xw,Gleiser:2010qt,Gleiser:2004an,Gleiserfeb2006,Gleiser:2009ys,Saffin,Amin:2010jq} and interactions~\cite{Hindmarsh:2007jb,Hindmarsh:2006ur}.
Several studies of the classical~\cite{Fodor:2008du,Fodor:2009kf,Salmi:2012ta,Gleiser:2008ty} and quantum~\cite{Hertzberg:2010yz,Kawasaki:2013awa,Saffin:2014yka} decay of these objects have also been performed.
A number of production methods have been studied: collapse of subcritical bubbles nucleated during first-order phase transitions~\cite{Copeland:1995fq},
production from collapsing domain wall networks~\cite{Hindmarsh:2007jb},
production from homogeneous field oscillations around a false vacuum minimum as a method of facilitating the formation of true vacuum bubbles~\cite{Gleiser:2003uu,Gleiser:2004iy,Gleiser:2007ts},
amplification of thermal fluctuations during inflation~\cite{FarhiGuth},
and production as a result of preheating at the end of inflation~\cite{Gleiser:2011xj,Amin:2010dc,Amin:2011hj,Amin:2010xe}.

In our study oscillons appear in two forms.
The first is the localized blobs of oscillating field in the planar background dynamics --- the planar equivalent of the oscillons in the corresponding one-dimensional field theory.
More interesting are the three-dimensional oscillons that form at the end of the fracturing of the walls.
The production mechanisms listed in the previous paragraph are based on the amplification of fluctuations around a homogeneous field background; 
thus the resulting oscillons are homogeneously distributed throughout the bulk.
In this sense, our mechanism is somewhat different since we have a strong localization along the collision axis.
Of course, the oscillons are still distributed uniformly in the directions transverse to the collision axis.

We now consider some simple oscillon properties of relevance to our domain wall collisions.
For simplicity, we only explicitly consider oscillons in the symmetric double-well.
First we demonstrate that an \emph{isolated} localized blob of field displaced from the minimum of the potential can indeed collapse to form an oscillon.
This problem has been studied using the assumption of exact spherical symmetry~\cite{Copeland:1995fq,Honda:2001xg}, and also for the case of nonspherical blobs in two spatial dimension~\cite{Adib:2002ff,Salmi:2012ta}.

As seen from our simulations, oscillons ultimately form from the fracturing of the walls and subsequent collapse of either a network of tubes or densely packed pockets with the field displaced from the true vacuum minimum in the interior.
As a very rudimentary approximation to this, we consider initial field profiles given by
\begin{equation}
  \phi_{init} = \phi_{true} + (\phi_{false}-\phi_{true})\exp\left(-\left(\frac{x^2}{a^2}+\frac{y^2}{b^2}+\frac{z^2}{c^2}\right)\right)
  \label{eqn:blob_profile}
\end{equation}
with $\dot{\phi} = 0$.
While this initial profile and the choice $\dot{\phi}=0$ are simplifications of what is seen in our simulations,
whether or not an oscillon forms cannot be too sensitive to the initial field configuration or we would not see them form at all.
In order to reduce the phase space of initial configurations, we further impose that $b^2=c^2$.
By choosing $a^2 > b^2$ we get cigar-like initial blobs and $a^2 < b^2$ gives us pancake like configurations.

\begin{figure}[t]
  \begin{center}
  \includegraphics[width=0.48\linewidth]{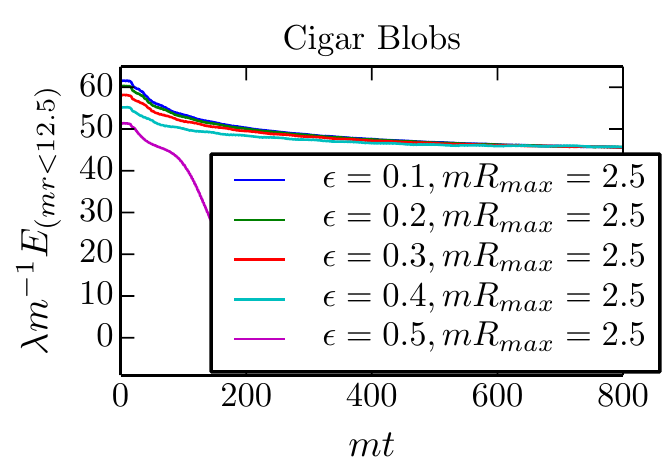}
  \includegraphics[width=0.48\linewidth]{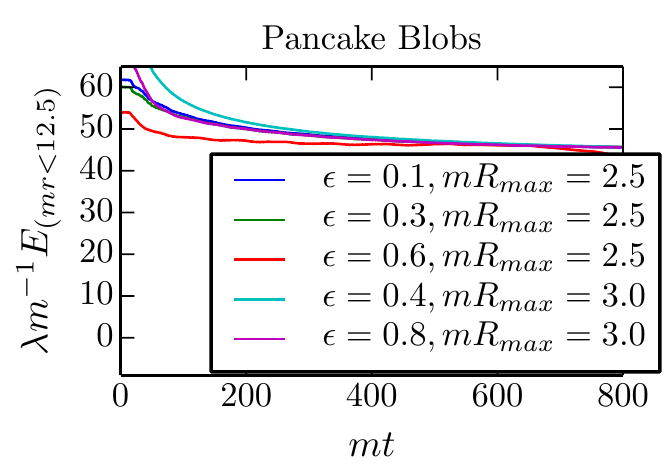}
  \end{center}
  \caption[Energy in a sphere of radius $12.5m^{-1}$ as a function of time for several choices of initial elliptical field profiles]{Energy in a sphere of radius $12.5m^{-1}$ as a function of time for several choices of initial elliptical field profiles given by~\eqref{eqn:blob_profile}.  We have parameterized the departure from spherical symmetry by $\epsilon^2 = 1-R_{min}^2/R_{max}^2$ where $R_{min} = \mathrm{min}(a,b)$ and $R_{max} = \mathrm{max}(a,b)$.  On the left we plot the results for cigar shapes and on the right for pancake shapes.  For definitions of the pancake and cigar configurations see the main text.}
  \label{fig:blob_energies}
\end{figure}

In~\figref{fig:blob_energies} we show the energy contained within a sphere of radius $12.5m^{-1}$ centered on the initial blob for several choices of the initial asymmetry.
Provided the blobs are not too asymmetrical, the energy within the sphere reaches a long-lived plateau indicating the presence of an oscillon.
For all initial conditions that result in an oscillon the plateau energy is the same, suggesting that the final oscillon states are all very similar for this particular model.\footnote{There exist models for which a range of different oscillons with different radii exist~\cite{Amin:2010jq}.}

We now provide some brief comments about the relationship between the 2D power spectrum after the walls have annihilated and the corresponding power spectrum for a single oscillon.
Since unbound forms of energy such as radiation escape the collision region, we can approximate the field near the collision site as
\begin{equation}
  \phi \approx \sum_{\alpha_i} \phi_{osc,i}({\bf x - x}_{\alpha_i},t)
\end{equation}
where the oscillons are located at positions ${\bf x}_{\alpha_i}$ and the profile of the $i$th oscillon is given by $\phi_{osc,i}$.
Similar expressions hold for other derived fields such as the energy density.
The distribution of $x_{\alpha}$ is determined by the (random) realization of the initial fluctuations, as well as the choice of initial planar background.
In the special case that $\phi_{osc,i}$ is independent of $i$ (\ie\ all of the oscillons have the same shape), the resulting power spectrum simplifies tremendously and we obtain
\begin{equation}
  \langle|\tilde{\phi}_k|^2\rangle = \left\langle\left|\sum_{\alpha_i}e^{i{\bf k}\cdot{\bf x}_{\alpha_i}}\right|^2\right\rangle\left\langle\left|\tilde{\phi}_{osc}(k)\right|^2\right\rangle  \equiv P_{form}(k)P_{oscillon}(k) \, .
  \label{eqn:spectrum_phasefac}
\end{equation}
In the general case,~\eqref{eqn:spectrum_phasefac} is replaced by a significantly more complicated expression, especially if there are many possible oscillon profiles and the positions and profiles are correlated.

Given the potential to extract information about the distribution of oscillons from measurements of two-point correlations, 
it is worthwhile to explore the possible spectra of individual oscillons as well as a characterize the range oscillon properties and time-dependence.
For our semiclassical simulations, we can extract this information directly in real space.
However, in numerical methods based on evaluation of a hierarchy of n-point correlation functions (such as the nPI formalism) this direct approach is not available.

\begin{figure}[t]
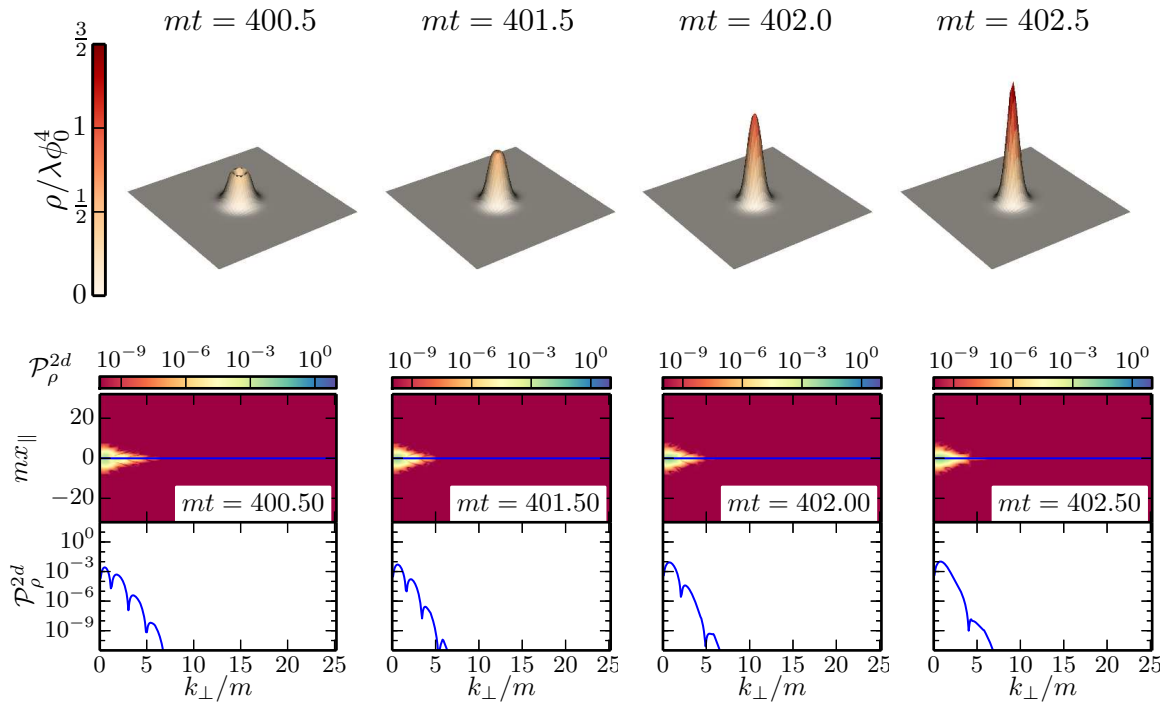

  \centering
  \includegraphics[width=0.99\linewidth]{{{oscillon_rhoslice_multipanel}}}\\
  \includegraphics[height=2in]{{{oscillon_pspec_04005}}} \hfill
  \includegraphics[height=2in]{{{oscillon_pspec_04015}}} \hfill
  \includegraphics[height=2in]{{{oscillon_pspec_04020}}} \hfill
  \includegraphics[height=2in]{{{oscillon_pspec_04025}}} 
  \caption[Evolution of an oscillon formed in the symmetric double-well potential]{A sequence of snapshots illustrating the time evolution of an oscillon formed from an initially spherical Gaussian field profile $\phi_{init} = 1-2e^{(r/r_{init})^2}$ with $mr_{init}=3$ in the symmetric double well.  In the top row are 2-dimensional slices of the energy density taken along a slice through the middle of the blob.  In the bottow row is the dimensionless 2-dimensional power spectrum as a function of position along the slicing axis (\emph{top panel, bottom row}) and also along the line indicated by the green line in the top panel (\emph{bottom panel, bottom row}).  An animation of the energy density is available at \oscillonrho\ and the power spectrum at \oscillonspec.}
  \label{fig:oscillon}
\end{figure}

A complete characterization of all oscillon properties in an arbitrary field theory is a rather daunting numerical task, so here we simply provide the spectra for a sample oscillon in the symmetric double-well potential.
The energy plateaus in~\figref{fig:blob_energies} provide some preliminary evidence that the oscillons themselves have a very narrow range of properties in this model, hence studying only a small sample may be sufficient.
\Figref{fig:oscillon} shows the energy density and 2D power spectrum for an oscillon formed from an initial spherical Gaussian field blob of radius $mr_{init}=3$.
After a short transient, the field quickly settles down into an oscillon configuration.
The energy density oscillates in time and looks like a shell of energy density that begins to expand outward before collapsing to form a sharp peak and subsequently expanding outward again.
Comparing the spectra with~\figref{fig:nondegen}, we see that the two isolated blobs of power away from the collision region in the asymmetric double-well collision are indeed oscillons that were ejected from the collision region.

\section{Summary of Mechanism}
\label{sec:summary_dynamics}
We explored the fully nonlinear three-dimensional evolution of colliding nearly planar domain wall-antiwall pairs for various choices of background solutions in the sine-Gordon and double-well potentials.
We now highlight the features of the dynamics that are most important in determining the qualitative outcomes of the collisions. Although the evolution of the fields is complex and difficult to intuit without the aid of numerical simulation,
once the results are known the qualitative behaviour is simple to understand.

Initially, the system is accurately described by a planar symmetric background $\phi_{bg}(x,t)$ and a collection of small nonplanar fluctuations $\delta\phi$.
The fluctuations couple to the background through the curvature of the potential evaluated on the background solution, $V''(\phi_{bg})$.
When the walls collide repeatedly or form oscillating bound states, this coupling drives an instability in the fluctuations which eventually pushes the entire system into the nonlinear regime.
Since the amplified fluctuations vary in the directions transverse to the collision, when they become large the walls develop noticable bumps and ripples.

\begin{figure}
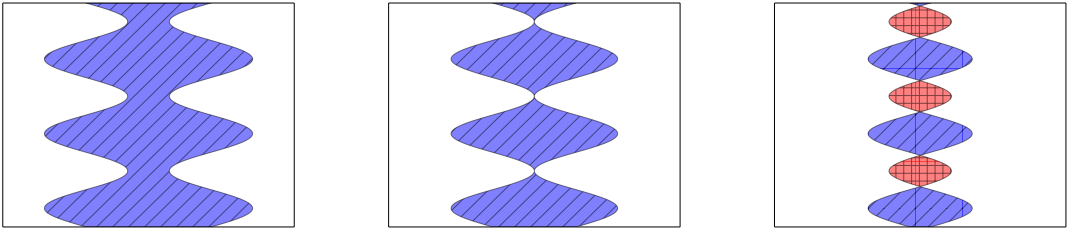

  \includegraphics[width=0.32\linewidth]{{{wall_cartoon_1}}}
  \includegraphics[width=0.32\linewidth]{{{wall_cartoon_2}}}
  \includegraphics[width=0.32\linewidth]{{{wall_cartoon_3}}}
  \caption[A heuristic view of the final collision between a pair of walls leading to the population of peaks in the field that subsequently condense to form oscillons]{A heuristic view of the final collision between a wall-antiwall pair.  White corresponds to the regions in the true vacuum, the blue angle hatched pattern to regions near the false vacuum minimum and the red hatched pattern to regions which have experienced a displacement due to the collision.  In the double well the field in the red region is initially displaced up the potential past the true vacuum, while in the sine-Gordon potential the field is displaced towards another minimum of the potential.  The red regions subsequently form the tubes of ``true vacuum'' that puncture the domain walls.
\emph{Left:} The two walls with just before the final collision with large ripples due to the pumping of transverse fluctuations in previous collisions.  
\emph{Center:} The first moments of the final collision.  Due to the large fluctuations in the location of the wall, the collision occurs asynchronously at different locations.  
\emph{Right:} The production of pockets of false vacuum as a result of the inhomogeneous nature of the final collision.}
  \label{fig:wall_cartoon}
\end{figure}

The details of the next stage of the evolution depend on the particular choice of planar background and potential.
In one scenario, which occurs for the sine-Gordon breathers with $v \geq 1$, the fluctuations ``pinch off'' to form an egg-carton like structure of field pockets.
The pockets of field are distributed in the plane orthogonal to the collision and centered on the collision location.
Outside the pockets the field is near the origin, while inside the field is displaced from the minimum towards a neighbouring vacuum.
In the second scenario, seen here in the case where the wall and antiwall repeatedly collide with each other, 
this stage is instead characterized by a final inhomogeneous collision between the wall-antiwall pair.
A heuristic illustration of the final collision is given in~\figref{fig:wall_cartoon}.
During this collision, punctures develop that thread the walls with regions of true vacuum.
These punctures then expand and eventually coalesce to leave behind an approximately two dimensional network of filaments.
Inside the filaments the field is near the false vacuum, while outside it is near the true vacuum.
From the viewpoint of observations restricted to an orthogonal slice of the field through the collision region the whole process resembles the nucleation, expansion and coalescence of a collection of two-dimensional bubbles.

In either of the two cases outlined above, the collision dynamics ultimately results in the formation of a highly inhomogenous field configuration with many peaks where the field is displaced from the true vacuum.
These peaks are distributed in the transverse plane, but they are localized along the collision axis.
As demonstrated in section~\ref{sec:oscillons}, \emph{isolated} peaks of the field can collapse nonlinearly to form oscillons.
When we have a field configuration with many peaks, we similarly expect some of them to form oscillons.
Of course, nonlinear interactions between the various peaks as the collapse is occurring leads to a more complicated scenario than the case of a single isolated peak,
but the effects of these interactions are insufficient to completely disrupt the production of oscillons.

\section{Conclusions}
\label{sec:conclusion}
We performed fully nonlinear three-dimensional simulations of parallel planar domain wall collisions, with the effects of initially small quantum fluctuations included.
This allowed us to probe nonlinear regimes inaccessible to the linear analysis of the fluctuations in~\cite{ref:bbm1}.
As anticipated from the resonances found in the linear theory, early in the evolution the fluctuations grow rapidly during collisions between the walls.
However, the most interesting phenomenology arises once the fluctuations begin to nonlinearly interact with each other, causing a complete breakdown of the original planar symmetry.
For the collisions we considered, this symmetry breakdown results in an extremely inhomogeneous dissolution of the walls
and eventually the production of oscillons distributed in the collision plane.
Our findings show a \emph{radical departure} from the behaviour in the symmetry reduced one-dimensional collisions.
The evolution is also completely different than the result if the backreaction of the fluctuations on the domain walls is treated as a homogeneous effect in the transverse plane,
such as a Hartree-like approximation would assume.
Therefore, we discovered a completely new phenomenology which can only be adequately studied using lattice simulations such as those used in this paper.

The dissolution of the walls is a consequence of our restriction to collisions between wall-antiwall pairs, which means there is no topological conservation law preventing the walls from eventually annihilating each other.
If we were to consider collisions between walls interpolating between different vacua as $|x| \to \infty$, then the post-collision state must contain domain wall like structures interpolating between the two different vacua at infinity.
Although the final dissolution of the walls will not occur in this case, our general finding that the planar symmetry is broken should continue to hold.
A specific example of this was our study of the fluctuations about a single isolated domain wall with an oscillating width.
In more general collisions, we expect that any walls remaining after the collision will not be produced with perfect planar symmetry, but instead will be bumpy.

Our results indicate a hurdle to constructing viable cosmological models predicated on the existence of many domain wall solutions.
In the absense of a topological charge preventing their annihilation, collisions lead to a mutual annihilation of the walls after only a few collisions.
This is much more rapid that the corresponding annihilation in the exactly planar one-dimensional theory.
Therefore, any braneworld model where we are confined to such a domain wall must take great care to ensure that these types of collisions are infrequent.

In~\cite{ref:bbm3} we consider the full nonlinear dynamics of SO(2,1) collisions between vacuum bubbles. We show that similar phenomenology arises in bubble collisions when the effects of fluctuations are included in the nonlinear problem.

\acknowledgments
We would like thank Andrei Frolov and Belle Helen Burgess for useful discussions and comments.
This work was supported by the National Science and Engineering Research Council of Canada and the Canadian Institute for Advanced Research.
JB was partially supported by the European Research Council under the European Community's Seventh Framework Programme (FP7/2007-2013) / ERC grant agreement no 306478-CosmicDawn.
Computations were performed on CITA's Sunnyvale cluster.

\bibliography{refs}{}
\bibliographystyle{JHEP}

\end{document}